\pdfoutput=1
\documentclass[
nofootinbib,
,amsfonts,amssymb,amsmath, twocolumn
]{revtex4-2}
\usepackage{fontawesome}

\usepackage{sastyle} 
\usepackage{amsmath,bm,mathtools,physics}
\usepackage[utf8]{inputenc}
\usepackage{bigints}

\numberwithin{equation}{section}

\usepackage{graphicx}
\usepackage{comment}
\graphicspath{ {images/} }

\usepackage{color}
\definecolor{darkblue}{rgb}{0.1,0.1,.7}
\usepackage[colorlinks, linkcolor=darkblue, citecolor=darkblue, urlcolor=darkblue, linktocpage]{hyperref} 

\newcommand{\tikzimage}[1]{\begin{aligned}
\includegraphics{#1.pdf}
\end{aligned}}

\def\jjk{\mathrm{JJK}}
\def\jkk{\mathrm{JKK}}
\def\jk{\mathrm{JK}}
\def\kk{\mathrm{KK}}

\def\ads{{\textrm{\tiny AdS}}}
\def\ds{{\textrm{\tiny dS}}}
\def\cft{{\textrm{\tiny CFT}}}

\newcommand*\JJskip{8mu}
\catcode`,\active
\newcommand*\JJ{\begingroup
	\catcode`\,\active
	\def ,{\mskip\JJskip\relax}%
	\doJJ
}
\catcode`\,12
\def\doJJ#1#2#3#4{%
	\Phi^{#1}_{#2}\biggl[\genfrac..{0pt}{}{#3}{#4}\biggr]%
	\endgroup
}
\allowdisplaybreaks
\begin{document}
%\title{Soft limits of spinning correlators in holographic and cosmological spacetime}
\title{Soft limits of gluon amplitudes in holography and cosmology}

%% Soft Limits of Gluons in cosmological and AdS spacetimes
\def\andname{}
\author{Soner Albayrak$^{\text{\faPencil}\,,\,\text{\faPencilSquare}\,,\,\text{\faPencilSquareO}}$}
\author{Savan Kharel$^{\text{\faEraser}}$}
\affiliation{$^{\text{\faPencil}}$Institute of Physics, University of Amsterdam, Amsterdam, 1098 XH, The Netherlands}
\affiliation{$^{\text{\faPencilSquare}}$Center for Theoretical Physics, National Taiwan University, Taipei 10617, Taiwan}
\affiliation{$^{\text{\faPencilSquareO}}$Department of Physics, Middle East Technical University, Ankara 06800, Turkey}
\affiliation{$^{\text{\faEraser}}$Department of Physics, University of Chicago, Chicago, IL 60637, USA}

\begin{abstract}
In this work, we extend the study of soft limits to (Anti) de Sitter spaces, investigating the analytic structure of holographic gluon correlators as part of a broader effort to reveal new symmetries and fundamental structures in gauge theories. By reorganizing perturbation theory in AdS to align with flat space unitarity, we analyze the contributions intrinsic to curved spacetime and their behavior in the soft limit. Our analysis uncovers schematic relations between \( (n+1) \)-point amplitude and \( n \)-point transition amplitudes in arbitrary-dimensional AdS, with explicit results derived for \( n=4 \) in \(\text{AdS}_{d+1}\).
\end{abstract}
\date{\today}
\maketitle
%\pacs{}
%\keywords{}
%\tableofcontents
\section{Introduction}
The study of scattering amplitudes has revealed interesting structures and patterns that showcase the deep symmetries inherent in quantum field theories (QFTs). Ranging from the streamlined form of MHV amplitudes to the geometric insight of the amplituhedron, these findings affirm the predictive power and internal consistency of QFTs. Theoretical developments like on-shell recursion relations and the double copy give us useful tools for decoding these structures, deepening our grasp of the foundational principles underlying particle interactions \cite{Parke:1986gb, Cachazo:2004kj, Britto:2004ap, Britto:2005fq, ArkaniHamed:2010kv}.

A time-honored tool in the study of scattering amplitudes is soft limits. These limits show a deep connection between the behavior of scattering amplitudes at low energies and the underlying symmetries of the theory. It is humbling to think that the idea of soft limits originated in quantum electrodynamics (QED) as early as 1937, with pioneering work in \cite{Bloch:1937pw}. Significant progress was made in 1958 by Low and collaborators \cite{Low:1954kd, GellMann:1954kc, Low:1958sn, kazes1959generalized, Yennie:1961ad}, and the formalism was later extended to gravitational interactions by Weinberg in 1965 \cite{Weinberg:1965nx}.

In particular, Weinberg's soft theorem is a prime example that illustrates the intricate relationship between low-energy phenomena and overarching symmetries. This  theorem demonstrates that in the soft limit, the scattering amplitude factorizes into a universal soft factor and the original amplitude, with the graviton removed. This result is independent of the details of the interaction; and, it imposes constraints on long-range forces, linking the behavior of low-energy gravitons to the underlying structure of quantum field theories \cite{Weinberg:1965nx}. 

Just when it seemed that soft limits and soft theorems, developed so many decades ago, looked ready to be shelved alongside other relics of theoretical physics in a dusty attic, it clawed its way back into relevance with insights that keep on giving. In recent years, Weinberg's soft theorem has been reframed as a Ward identity of Bondi–van der Burg–Metzner–Sachs (BMS) symmetry \cite{Strominger:2013jfa}. Strominger proposed that the first subleading soft terms arise as consequences of BMS symmetry, a claim validated at tree level by Cachazo and Strominger, who also demonstrated the universality of the second-order subleading corrections \cite{Cachazo:2014fwa}. Soon after, analogous subleading corrections were established within Yang–Mills theory \cite{Casali:2014xpa}. This progress has highlighted that soft gluons are essential for uncovering fundamental structures in gauge theories, shedding new light on symmetries, factorization, and infrared behavior in flat space.

In this paper, we similarly investigate the behavior of amplitudes in Yang–Mills theory under a soft expansion, but in a curved space context. Examining how the intricate structure of soft theorems extends to non-flat backgrounds could potentially reveal new features and simplify computations across dimensions. AdS correlators, for instance, offer a generalized structure that encompasses scattering amplitudes obtained in the flat space limit, with seemingly analogous features such as recursion relations and the double copy emerging in cosmological wavefunctions as well \cite{Albayrak:2020fyp, Armstrong:2020woi, Zhou:2021gnu, Cheung:2022pdk}. Indeed, the study of soft limits in AdS may uncover rich structures that could advance our understanding of both QFTs and holography. In a related cosmological context, Maldacena's soft theorem imposes a nontrivial consistency condition on single-clock inflationary models \cite{Maldacena:2002vr, Creminelli:2004yq}. This foundational work has inspired various generalizations in the inflationary setting \cite{Creminelli:2012ed, Assassi:2012zq, Hinterbichler:2012nm, Flauger:2013hra, Pimentel:2013gza, Goldberger:2013rsa, Hinterbichler:2013dpa, Joyce:2014aqa}: such developments in cosmology and AdS computations are complementary, as it is now well established—though perhaps underappreciated—that AdS correlators are directly linked to cosmological wavefunction coefficients (see Appendix \ref{wave}).

%and we analyze here the soft limits of gluons in AdS across arbitrary dimensions, specifically examining how the correlator behaves under this limit. with all insights here equally relevant to cosmological wavefunctions.

This paper is organized as follows. Section~\ref{sec:preliminaries}  covers background on scattering amplitudes and soft limits in both flat and curved spacetimes. We also review the AdS perturbation theory in momentum space, which we choose as our \emph{modus operandi}.\footnote{Naturally, the momentum space formalism offers the most intuitive framework for probing soft limits; and fortunately, it has also proven to be a rich and productive arena for the study of (A)dS correlators over the last decade \cite{Raju:2012zr, Albayrak:2019asr, Albayrak:2020bso, Albayrak:2020isk, Albayrak:2020fyp, Albayrak:2023kfk, Baumann:2019oyu, Baumann:2019oyu, Baumann:2020dch, Baumann:2021fxj, 
Bzowski:2013sza, Bzowski:2015pba, Bzowski:2018fql, Bzowski:2019kwd, Bzowski:2020kfw, Bzowski:2022rlz, Bzowski:2023jwt, Jain:2020rmw, Jain:2021vrv, Jain:2023idr, Isono:2018rrb, Isono:2019wex}}. We then analyze
the four-point AdS amplitude's soft limit in Section~\ref{sec:soft_limit_4pt}, deriving explicit expressions that recast these amplitudes using lower-point transition amplitudes. Section~\ref{sec:higher_point} extends this soft analysis to higher-point amplitudes, demonstrating that the leading soft behavior of the $(n+1)$-point amplitude relates to products of $n$-point transition amplitudes. Finally,  Section~\ref{sec:conclusion} summarizes results and discusses future directions in soft limits and holography in curved spacetime.

Technical derivations are provided in the appendices. Appendix~\ref{wave} reviews wavefunction coefficients and boundary correlators in (A)dS; Appendix~\ref{appendix unitarity} discusses unitarity and propagator decomposition; and, Appendix~\ref{appendix bessels} details Bessel function integrals relevant to our soft limit analysis. Further explicit results are provided in  Appendix~\ref{sec:ads4three}.

\emph{Note: While this work was nearing completion, \cite{Chowdhury:2024wwe} appeared which partially overlaps with our results.}

\section{Preliminaries}
\label{sec:preliminaries} 

\subsection{Soft limit in flat and curved spaces}
\label{sec:soft limit preliminary} 

Consider an \mbox{$(n+1)-$}point scattering amplitude $\cA(\bm{q},\bk_1^{h_1},\dots, \bk_n^{h_n})$ in flat space, where $\bm{q}$ \& $\bk_i$ are momenta of the external legs, $h_i$ run over all physical states, and we will take $\bm{q}$ to be small. This amplitude can be written as a sum of infinitely many Feynman diagrams, where each Feynman diagram is of the form $\bigints (\dots)\left[\prod\limits_{j\in\text{internal legs}}(P_j^2-m^2)\right]^{-1}\left[\prod\limits_{i\in\text{loops}}d^dl_i \right]$ for internal leg momentum norm $P_j$. We see that all Feynman diagrams are non-singular in the soft limit \emph{except} those for which $\bm{P}=\bm{q}+\bm{k}_i$, as $1/\left(P_j^2-m^2\right)\sim 1/\bm{q}\.\bm{k}_i$. But these are precisely the diagrams in which the soft leg combines with a single hard leg through a three-point vertex. We can then go ahead and collect all diagrams with same $\bm{k}_i$ and sum over them, giving us the full $n-$point amplitude; thus, we arrived at the well-known result in flat space: \emph{$(n+1)-$point amplitudes are proportional to $n-$point amplitudes in the soft limit}.

%%SK We may want to rephrase this a bit. But its a minor point and needs discussion

Let us illustrate this point further in gauge theories. The propagator of a gauge field in flat space is\footnote{We are suppressing the color factors throughout this paper.}
\be 
\label{propagator}
\Pi_{\mu\nu}(\bk)=\frac{\sum\limits_{h=1}^2\e_\mu^h(\bk)\e_{\nu}^{h^*}(\bk)}{\bk^2-m^2+i\e}
\ee 
for the polarization vectors $\e^\mu$,\footnote{A helicity $h^*$ is a shorthand notation such that $\e_\mu^{h^*}(\bk)\coloneqq\left(\e_\mu^h(\bk)\right)^*$.} which then means the color ordered partial amplitude $\cA_{n+1}$ reads as
\begin{multline}
	\cA_{n+1}\left(\bm{q}^{h_q},\bk_1^{h_1},\dots,\bk_n^{h_n}\right)=\sum\limits_{i=1}^n\sum\limits_{h_0=\pm}\frac{\cA_{3}\left(\bm{q}^{h_q},\bk_i^{h_i},\bk_0^{-h_0}\right)}{\bk_0^2-m^2+i\e}
	\\\x\cA_n\left(\bk_0^{h_0},\bk_1^{h_1},\dots,\bk_{i-1}^{h_{i-1}},\bk_{i+1}^{h_{i+1}},\dots,\bk_n^{h_n}\right)+\dots
\end{multline}
for $\bk_0\coloneqq\bm{q}+\bk_i$, where the other terms are higher order in small $\bm{q}$.

This nice form has nothing to do with the properties of gauge field; in fact, it is a consequence of unitarity in flat space.\footnote{See Appendix~\ref{appendix unitarity} for a brief discussion of this point.} Indeed, the numerator of a propagator \emph{for any spin} must be equal to the sum over physical spin states, hence the internal line of momentum $\bk_0$ practically splits as two external states, allowing the factorization of the full expression in terms of lower point amplitudes.

When we move beyond the flat space, the implications of unitarity is no longer as clear: indeed, we will see by an explicit example that the AdS amplitude\footnote{By \emph{amplitudes} in AdS, we refer to the bulk duals of boundary CFT correlators.} actually factorizes into a lower point \emph{transition amplitude}, not a \emph{vacuum amplitude} (we'll clarify our terminology below).

Another complication that arises in AdS is the practicality of the soft limit: as we quickly reviewed above, the soft limit is particularly useful as some Feynman diagrams are dominant to the others in this limit, allowing us to rewrite the amplitude in terms of a lower-point one.\footnote{One actually does not need the perturbative approach to show this; in fact, it is conceptually far more satisfying to derive it via a BCFW deformation. Nevertheless, diagrammatic approach is more familiar and is sufficient to make our points both in flat space and (A)dS.} Unlike their flat space counterparts, none of the propagators in an AdS diagram develops a dominant singularity in soft limit, hence the flat-space arguments do not immediately carry over. Nevertheless, it is useful and important to classify different Witten diagrams and investigate if and how flat space limit and soft limit commutes.

Let us expand on this point further. We know that gluon and graviton tree-level Feynman diagrams can be brought to the form
\bea[eq: tree leven gluon and graviton Witten diagrams]
W^{\text{Tree}}_{\text{gluon}}\sim &
\cD^{m,n,r}_\text{gluon}
\int_0^\infty
\frac{ dz_1}{z_1^{d+1}}\dots \frac{ dz_{r+s}}{z_{r+s}^{d+1}}
\left(\prod\limits_{c=1}^rz_c^4\right)
\nn\\&\hspace*{-2em}\x\left(
\prod\limits_{a=1}^{m}
\f^{d-2}_{d-2}(k_a,\tl z_a)
\right)
\left(
\prod\limits_{b=1}^{n}
\int
p_b dp_b
\JJ{d-2}{d-2}{q_b,p_b}{\widehat z_{2b-1},\widehat z_{2b}}
\right)
\\
W^{\text{Tree}}_{\text{graviton}}\sim &
\cD^{m,n,r}_\text{graviton}
\int_0^\infty
\frac{ dz_1}{z_1^{d+1}}\dots \frac{ dz_{r}}{z_{r}^{d+1}}
\left(\prod\limits_{c=1}^rz_c^8\right)
\nn\\&\hspace*{-2em}\x\left(
\prod\limits_{a=1}^{m}
\f^{d-4}_{d}(k_a,\tl z_a)
\right)
\left(
\prod\limits_{b=1}^{n}
\int
p_b dp_b
\JJ{d-4}{d}{q_b,p_b}{\widehat z_{2b-1},\widehat z_{2b}}
\right)
\eea
where $m$, $n$, $r$, and $s$ denote the number of external legs, internal legs, three point vertices, and four point vertices respectively. Here $\cD$ and $\Phi$ denote some differential operators and some functions whose details will not be important \cite{Albayrak:2020bso}; what is important is that they are both non-singular in the soft limit and the only-singularity in these Witten diagrams come through the term $\phi$ which is simply a modified Bessel function of the second kind up to a normalization. This is why unlike their flat-space counterparts all Feynman diagrams in AdS have the similar behavior in the soft-limit, precluding a clear relation between $(n+1)-$point and $n-$point AdS amplitudes.

\subsection{A short review of perturbation theory in (A)dS for Yang Mills theory}
\label{sec:ads_preliminaries}

Let's look at Yang-Mills equations in AdS. In this paper, we adopt conventions aligned with prior work in the field \cite{Raju:2012zr, Albayrak:2018tam, Albayrak:2019yve};\footnote{Alternatively, see \cite{Paulos:2011ie, Chu:2023pea, Chu:2023kpe} for perturbation theory of AdS Yang-Mills theory as \emph{Mellin amplitudes} instead.} however, we review these conventions here for completeness and introduce a diagrammatic notation for the reader.
We consider AdS$_{d+1}$ whose metric in Poincar\'e patch reads as $ ds^2=z^{-2}(dz^2+\eta_{ij}dx^i dx^j)$. In the axial gauge, the bulk to boundary propagator for a spacelike momentum $\bk$ reads as\footnote{For simplicity, we will focus on spacelike external momenta, though our results can be translated to other cases by taking care of appropriate transformations (e.g. using Hankel function of the first kind instead of modified Bessel function of the second kind, and so on) \cite{Raju:2012zs}.}
\be 
\hspace*{-.6em}A_i^{h_a}(\bk_a,z)\coloneqq\e_i^{h_a}(\bk_a)k^\nu z^\nu K_\nu(k_az)\doteq\tikzimage{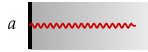}
\ee 
where $h_a$ is the helicity of the polarization and $k_a$ is the norm of momentum $\bk_a$, i.e. $k_a^2=\bk_a\.\bk_a$.\footnote{
This form of the propagator is dictated by regularity in the bulk and the compatibility with the relevant Ward identity up to an immaterial overall normalization \cite{Albayrak:2023kfk}.
} Here, we define $\nu\equiv d/2-1$ for brevity. We also define the \emph{normalizable} mode as
\be 
\hat A_i^{h}(\bk,z)\coloneqq\e_i^{h}(\bk)k^\nu z^\nu J_\nu(kz)\doteq\tikzimage{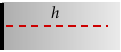}
\ee 
which is intimately tied to the transition amplitudes.\footnote{Transition amplitudes are bulk duals of CFT correlators between states; more precisely, if we start with a CFT vacuum correlator dual to an ordinary AdS amplitude and revert some of the external modes of the amplitude from non-normalizable to normalizable (hence converting the amplitude to a transition amplitude), then the dual CFT correlator gets transformed into a new one which measures the probability of transition between two coherent states \cite{Raju:2011mp}.} We also define an \emph{off-shell} version of this normalizable mode
\be 
\hat A_i^{h}(\bk,p,z)\coloneqq\e_i^{h}(\bk) k^\nu z^\nu J_\nu(pz)\doteq\tikzimage{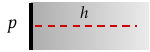}
\ee 
which plays a direct role in the decomposition of bulk-to-bulk propagator.\footnote{
	If we consider an abstract $d+1$ dimensional flat space, we could embed our $d-$dimensional momentum $\bk_i$ with the parameter $p$ as $\mathbb{k}=(\bk_i,p)$. In the flat space limit, we have $p\rightarrow -i k$, which turns $\mathbb{k}$ into a null $d+1-$dimensional momentum that parametrizes the scattering amplitude that lives in the flat space limit. Note that momentum conservation for $\mathbb{k}$ requires sum of the norms $k$ to be zero, which is precisely the flat space limit in this formalism. Therefore, we will call any leg with a generic parameter $p$ \emph{off-shell}, as the corresponding $\mathbb{k}^2\ne 0$. Note that this usage is consistent with the fact that the denominator of the bulk to bulk propagator diverges in the on-shell limit. 
} In axial gauge, the propagator is 
\be 
\label{eq: propgagtor}
 G_{ij}(\bk,z,z')\coloneq\int\limits_{\R^+}\frac{pdp}{2}\frac{\left[(zz')^\nu J_\nu(pz)J_\nu(pz')\right]T_{ij}}{k^2+p^2+i\e}\doteq\tikzimage{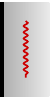}
\ee 
for the tensor structure
\be 
T_{ij}=\eta_{ij}+\frac{\bk_i\bk_j}{p^2}
\ee 
One can show that the propagator can be written down in a form
\be 
\label{eq: decomposition of AdS propagator}
G_{ij}(\bk,z,z')=G^L_{ij}(\bk,z,z')+\sum\limits_h\int\limits_{\R^+}\frac{pdp}{2k^{2\nu} }\frac{\hat A^{h^*}_i(\bk,p,z)\hat A^{h}_j(\bk,p,z')}{k^2+p^2+i\e}
\ee 
which in our diagrammatic notation means
\be 
\label{eq:propagator decomposition}
\tikzimage{propagator}\,=\,\tikzimage{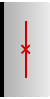}\,+\,\sum\limits_h\bigintssss\limits_{\R^+}\frac{pdp}{2k^{2\nu}(k^2+p^2)}\,\tikzimage{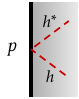}
\ee 
where we define the \emph{longitudinal} part of the bulk-to bulk propagator as
\be 
G^L_{ij}(\bk,z,z')=\frac{\bk_i\bk_j}{4\nu k^2}\left(\frac{\min\left(z,z'\right)}{\max\left(z,z'\right)}\right)^{\nu}\doteq\tikzimage{propagatorL}
\ee 
In the flat space limit \cite{Raju:2012zr}, this part disappears in accordance with the unitarity; thus, any diagram that contains a cross is inherently a curved space contribution.\footnote{As we discuss in Appendix~\ref{appendix unitarity}, unitarity ensures that the flat space propagator takes the form in \equref{propagator}, i.e. its numerator is a sum of \emph{physical degrees of freedom} alone. Because any diagram with a crossed leg contains a longitudinal degree of freedom that is absent in that sum, all such diagrams have to vanish in the flat space limit.}

We have omitted the momentum $\bk$ that flows in the legs in the above diagrams; in the rest of the paper, any momentum that can be fixed by momentum conservation will be omitted in the diagrams. Likewise, we will assume that any explicit helicity $h_i$ is summed over and any explicit momentum $p_i$ are integrated against the measure $\frac{p}{2k^{2\nu}(k^2+p^2)}$, in line with \equref{eq:propagator decomposition}. For instance, the decomposition of $4-$point $s-$channel Witten diagram would look like
\be
\label{eq: four point decomposition}
\tikzimage{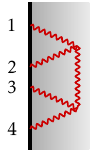}\,=\,\tikzimage{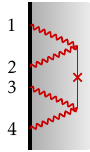}\,+\,\tikzimage{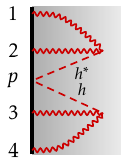}
\ee

%%%%

\section{Soft limit of four point amplitude}
\label{sec:soft_limit_4pt}

Let us start the discussion of the soft limit in (A)dS with the analysis of the simplest non-trivial case, i.e. soft limit of four point amplitudes. The actual computation of soft limit of a Witten diagram can be approached with three different methods:
\begin{enumerate}
	\item Direct computation
	\item Using the differential representation in \equref{eq: tree leven gluon and graviton Witten diagrams}
	\item Utilizing the \emph{``unitary''} decomposition of the propagator in \equref{eq:propagator decomposition}
\end{enumerate}
We will carry out all of these approaches to investigate their pros and cons. For concreteness, we will consider the $s-$channel computation.

\subsection{Direct approach}
The expressions for the respective Witten diagram reads 
\be 
W=\int\limits_0^\infty \frac{dz_L}{z_L^{d-3}}\frac{dz_R}{z_R^{d-3}}A_i^{h_1}(\bk_1,z_L)A_j^{h_2}(\bk_2,z_L)A_k^{h_3}(\bk_3,z_R)A_l^{h_4}(\bk_4,z_R)
\\\x V^{ijm}(\bk_1,\bk_2,-\bk_1-\bk_2)G_{mn}(\bk_1+\bk_2,z_L,z_R)V^{kln}(\bk_3,\bk_4,\bk_1+\bk_2)
\ee 
for the three point vertex factor
\begin{multline}
V^{\mu\nu\rho}(\bm{k}_1,\bm{k}_2,\bm{k}_3)\equiv{}
\frac{i}{\sqrt{2}}
\Big(
\eta^{\mu\nu}(\bm{k}_1-\bm{k}_2)^\rho\\+\eta^{\nu\rho}(\bm{k}_2-\bm{k}_3)^\mu+\eta^{\rho\mu}(\bm{k}_3-\bm{k}_1)^\nu\Big)
\end{multline}
One can immediately see that the expression becomes
\be 
\frac{W}{k_1^\nu k_2^\nu k_3^\nu k_4^\nu}=&S_1
\int\limits_{\R^+}\frac{pdp}{2}\frac{\jkk_{2+\nu,\nu}(p,k_1,k_2)\jkk_{2+\nu,\nu}(p,k_3,k_4)}{q^2+p^2+i\e}
\\&+S_2
\int\limits_{\R^+}\frac{dp}{2p}\frac{\jkk_{2+\nu,\nu}(p,k_1,k_2)\jkk_{2+\nu,\nu}(p,k_3,k_4)}{q^2+p^2+i\e}
\ee  
for $\bq=\bk_1+\bk_2$, where $S_i$ are some polarization structures\footnote{
For brevity, we define
\bea 
S_1=\left[\e_i^{h_1}(\bk_1)\e_j^{h_2}(\bk_2)V^{ijm}(\bk_1,\bk_2,-\bq)\right]\eta_{mn}\left[\e_k^{h_3}(\bk_3)\e_l^{h_4}(\bk_4)V^{kln}(\bk_3,\bk_4,\bq)\right]\\
S_2=\left[\e_i^{h_1}(\bk_1)\e_j^{h_2}(\bk_2)V^{ijm}(\bk_1,\bk_2,-\bq)\right]\bq_m\bq_n\left[\e_k^{h_3}(\bk_3)\e_l^{h_4}(\bk_4)V^{kln}(\bk_3,\bk_4,\bq)\right]
\eea 
} and $\jkk$ is a known combination of Appell’s hypergeometric functions as detailed in Appendix~\ref{appendix bessels}.

As far as the authors are aware, closed formulae for the symbolic integration of products of Appell's function have not been made available in the literature; however, soft limit of these expressions are more easily trackable; for instance, for $\bk_1\rightarrow 0$, \equref{eq: soft limit of JKK and JJK} indicates that\footnote{We also used the nice identity
\be 
P_{-1}^{-\nu}(x)=\frac{(1-x)^{\nu /2} (x+1)^{-\nu /2}}{\Gamma (\nu +1)}\quad\text{for }\nu\in\N/2
\ee 
}
\be
\lim\limits_{\bk_1\rightarrow 0}\jkk_{2+\nu,\nu}(p,k_1,k_2)=k_1^{-\nu}\left(\frac{\Gamma (\nu )}{2^{1-\nu }}\frac{p^\nu k_2^{-\nu}}{p^2+k_2^2}+\cO(k_1)\right)
\ee
with which we observe
\be 
\lim\limits_{\bk_1\rightarrow 0}W\sim
\int\limits_{\R^+}\frac{\# p^{\nu+1}+\# p^{\nu-1}}{\left(k_2^2+p^2\right)^2}\jkk_{2+\nu,\nu}(p,k_3,k_4)dp
\ee  
where we remind the reader that $\nu=d/2-1$. With $\jkk$ being the scalar part of a three point transition amplitude,\footnote{\label{footnote T}The three point amplitude reads as
\be
T^{h_1h_2h_3}_{\bk_1,\bk_2,\bk_3}=V^{ijk}(\bk_1,\bk_2,\bk_3)\int\frac{dz}{z^{d+1}}
A_i^{h_1}(\bk_1,z)A_j^{h_2}(\bk_2,z) A^{h_3}_k(\bk_3,z)
\ee
whereas the \emph{transition} amplitude becomes
\be
	T^{h_1h_2h}_{\bk_1,\bk_2;\bq,p}=V^{ijk}(\bk_1,\bk_2,\bq)\int\frac{dz}{z^{d+1}}
	A_i^{h_1}(\bk_1,z)A_j^{h_2}(\bk_2,z)\hat A^{h}_k(\bq,p,z)
\ee} this form suggests a connection between $n-$point amplitudes and $(n-1)-$point \emph{transition} amplitudes, a connection that will be clearer below.

\subsection{Differential Representation}
We presented the schematic form of the tree-level Witten diagrams in \equref{eq: tree leven gluon and graviton Witten diagrams}; for the respective Witten diagram at hand, the relevant formula becomes
\begin{multline}
W=\cD^{ijkl}\int\limits_0^\infty \frac{dz_L}{z_L^{d-3}}\frac{dz_R}{z_R^{d-3}}\int\limits_{\R^+}\frac{p dp }{2}A_i^{h_1}(\bk_1,z_L)A_j^{h_2}(\bk_2,z_L)\\\x \frac{(z_L z_R)^{\nu}J_\nu(pz_L)J_\nu(pz_R)}{q^2+p^2-i\e}A_k^{h_3}(\bk_3,z_R)A_l^{h_4}(\bk_4,z_R)
\end{multline}
for the operator
\begin{multline}
\cD^{ijkl}=V^{ijm}(\bk_1,\bk_2,-\bk_1-\bk_2)V^{kln}(\bk_3,\bk_4,\bk_1+\bk_2)\\\x\left(\frac{\eta_{mn}q^2-\bq_m \bq_n}{i q^2}+\frac{\bq_m \bq_n}{iq^2}\lim\limits_{q\rightarrow0}\right)
\end{multline} 
It is then straightforward to carry out $z_L$ and $z_R$ integrations to obtain the form
\begin{multline}
W=\left[\e_i^{h_1}(\bk_1)\e_j^{h_2}(\bk_2)\e_k^{h_3}(\bk_3)\e_l^{h_4}(\bk_4)\right]k_1^\nu k_2^\nu k_3^\nu k_4^\nu\\\x\cD^{ijkl}
\int\limits_{\R^+}\frac{pdp}{2}\frac{\jkk_{2+\nu,\nu}(p,k_1,k_2)\jkk_{2+\nu,\nu}(p,k_3,k_4)}{q^2+p^2+i\e}
\end{multline}

This short exercise emphasize the main advantage of the differential representation as argued in \cite{Albayrak:2020bso}: by relating various terms among each other, the computations reduce to a single integration whereas everything else is a simple application of limits and algebra, all of which can be rather efficiently automated by a software program such as \texttt{Mathematica}.\footnote{One might object that integrations can also be carried out in Mathematica hence, they would claim, there is not a real simplification. The problem is that symbolic integrations \emph{can not} be parallelized in \texttt{Mathematica} (nor in \texttt{Maple} or similar symbolic computation environments.) whereas algebraic computations can be efficiently conducted on multiple threads. Compounding the fact that arbitrary spacetime dimensional curved space computations are rather heavy on the CPU with the observation that the main progress in processing units for the past decade has been rather on increasing their numbers on a socket than increasing their single-thread power, we can see why trading symbolic integration for symbolic algebra is so computationally desirable.} In our particular focus of soft limit in this paper, this advantage still holds its merit as the soft limit commutes with the relevant operators in this representation; indeed, just as the previous case, we are left with a form
\be 
\lim\limits_{\bk_1\rightarrow 0}W\sim
\cD\int\limits_{\R^+}\frac{p^{\nu+1}}{\left(k_2^2+p^2\right)^2}\jkk_{2+\nu,\nu}(p,k_3,k_4)dp
\ee 
which still suggests a connection between the soft limit of the four point amplitude and the three point transition amplitude. 

Even though this approach seems to be less computationally intensive compared to the direct approach, it is too opaque to generate a physical intuition for the soft limit in (A)dS, so we next turn to a different representation for the same expression.

\subsection{``Unitary'' decomposition}
We discuss in Appendix~\ref{appendix unitarity} how the unitarity imposes stringent conditions on the propagator in flat space and how we can try to extrapolate that knowledge into homogeneous spaces. We show how this could be diagrammatically represented in \equref{eq: four point decomposition}; in equations, it reads as
\be 
W=&V^{ijm}(\bk_1,\bk_2,-\bq)V^{kln}(\bk_3,\bk_4,\bq)\int\frac{dz_Ldz_R}{z_L^{d+1}z_R^{d+1}}G^L_{mn}(q,z_L,z_R)
\\&\x A_i^{h_1}(k_1,z_L)
A_j^{h_2}(k_2,z_L) A_k^{h_3}(k_3,z_R)
A_l^{h_4}(k_4,z_R)
\\+&
V^{ijm}(\bk_1,\bk_2,-\bq)V^{kln}(\bk_3,\bk_4,\bq)\sum\limits_h\int\limits_{\R^+}\frac{pdp}{2q^{2\nu}(q^2+p^2+i\e)}\\&\x\int\frac{dz_L}{z_L^{d+1}}A_i^{h_1}(k_1,z_L)A_j^{h_2}(k_2,z_L)\hat A^{h^*}_m(\bq,p,z_L)
\\&\x\int\frac{dz_R}{z_R^{d+1}}
A_k^{h_3}(k_3,z_R)A_l^{h_4}(k_4,z_R)\hat A^{h}_n(\bq,p,z_R)
\ee 
for $\bq=\bk_1+\bk_2$ as before. We can massage this equation into
\begin{multline}
\label{eq: s channel witten diagram in unitary decomposition}
W=\sum\limits_h\int\limits_{\R^+}\frac{pdp}{2q^{2\nu}(q^2+p^2+i\e)}T^{h_1h_2h^*}_{\bk_1,\bk_2;\bq,p}T^{h_3h_4h}_{\bk_3,\bk_4;-\bq,p}
\\+\frac{S_2 k_1^\nu k_2^\nu k_3^\nu k_4^\nu}{4\nu q^2}\Bigg(\int\limits_{0}^{\infty}\frac{dz_R}{z_R^{\nu+3}}K_\nu(k_3z_R)K_\nu(k_4z_R)\kk^{(1)}_{\nu-2,\nu}(k_1,k_2,z_R)
\\+\int\limits_{0}^{\infty}\frac{dz_R}{z_R^{-\nu+3}}K_\nu(k_3z_R)K_\nu(k_4z_R)\kk^{(2)}_{-\nu-2,\nu}(k_1,k_2,z_R)\Bigg)
\end{multline}
where  $T$'s are defined in the footnote~\ref{footnote T} and where the physical flat-space degrees of freedom manifestly appear in the first term through transverse polarizations and inherently-(A)dS contribution in the second piece carries the longitudinal polarization. If we now take the soft limit \mbox{$\bk_1\rightarrow 0$}, this becomes
{\small\begin{multline}
\label{eq: soft limit in unitary decomposition for four point}
\lim\limits_{\bk_1\rightarrow 0} W=\sum\limits_h\int\limits_{\R^+}\frac{pdp}{2k_2^{2\nu}(k_2^2+p^2+i\e)}T^{h_3h_4h}_{\bk_3,\bk_4;-\bq,p}\left(\lim\limits_{\bk_1\rightarrow 0} T^{h_1h_2h^*}_{\bk_1,\bk_2;\bq,p}\right)
\\+\frac{S_2\Gamma (\nu )k_2^\nu k_3^\nu k_4^\nu}{32\nu k_2^{\nu+2}}\int\limits_{0}^{\infty}dz_RK_\nu(k_3z_R)K_\nu(k_4z_R)\Bigg(\frac{z_R^{\nu-3}\Gamma
	\left(-\frac{\nu+2}{2}\right)  \Gamma
	\left(-\frac{3\nu+2}{2}\right)}{2^{\nu+2}k_2^{-3 \nu-2}}
\\+\frac{2\Gamma (-\nu )}{z_R^{5}k_2^{-2 \nu }}\bigg[\frac{ \, _1F_2\left(\frac{\nu-2
	}{2};\frac{\nu}{2},\nu +1;\frac{k_2^2
		z_R^2}{4}\right)}{\nu-2 }
+\frac{\, _1F_2\left(\frac{-\nu-2
	}{2};-\frac{\nu}{2},\nu +1;\frac{k_2^2
		z_R^2}{4}\right)}{\nu+2 }
\bigg]
\\
-\frac{\Gamma (\nu )}{z_R^{2\nu+5}2^{-2\nu-1}}\bigg[
\frac{ \,
	_1F_2\left(-\frac{\nu+2 }{2};1-\nu ,-\frac{\nu}{2};\frac{k_2^2 z_R^2}{4}\right)}{\nu+2}
\\	
+\frac{ \, _1F_2\left(-\frac{3\nu+2
	}{2};1-\nu ,-\frac{3\nu}{2};\frac{k_2^2
		z_R^2}{4}\right)}{3 \nu+2 }	
\bigg]\Bigg)
\end{multline}\normalsize
}where we use equation~\eqref{eq: limits of kk, jk, etc}.\footnote{Here, we are commuting the integrations with the soft limit which is not necessarily warranted. However, as the AdS amplitude in soft limit is towards a regular point of the amplitude (i.e. it is non-singular in the limit), there is no reason to expect it to develop any non-analyticity as long as the integral is convergent.   The integral \emph{is} non-convergent in some cases, which simply signals the necessity of a regularization scheme, and one can argue that the commutation of the limit with the integration (even after the regularization) is not warranted in those cases. In such a situation, one can start with non-physical kinematic regimes with convergent integral expressions, and investigate physical configurations through careful analytic continuations. It is of interest what such an analysis would reveal; however, it is beyond the scope of this paper.} The computation of the soft limit of a transition three point amplitude is relatively straightforward and it reads as
\be 
\label{eq: soft limit of transition amplitude}
\lim\limits_{\bk_1\rightarrow 0} T^{h_1h_2h^*}_{\bk_1,\bk_2;\bq,p}%=& \lim\limits_{\bk_1\rightarrow 0} k_1^\nu V^{ijm}(\bk_1,\bk_2,-\bq)\int\frac{dz_L}{z_L^{d+1}}A_i^{h_1}(k_1,z_L)A_j^{h_2}(k_2,z_L)\hat A^{h^*}_m(\bq,p,z_L)
%\\
%=&\lim\limits_{\bk_1\rightarrow 0} k_1^\nu V^{h_1h_2h^*}(\bk_1,\bk_2,-\bq)\int\frac{dz_L}{z_L^{d+1-3\nu}}K_\nu(k_1 z_L)K_\nu(k_2 z_L)J_\nu(p z_L)
%\\
%=&\lim\limits_{\bk_1\rightarrow 0} k_1^\nu V^{h^*h_1h_2}(-\bq,\bk_1,\bk_2)\int\frac{dz_L}{z_L^{d+1-3\nu}}K_\nu(k_1 z_L)K_\nu(k_2 z_L)J_\nu(p z_L)
%\\
%=&-i\sqrt{2}[-\bq\.\e^{h_1}(\bk_1)][\e^{h^*}(-\bq)\.\e^{h_2}(\bk_2)]\lim\limits_{\bk_1\rightarrow 0} k_1^\nu 
%\int\frac{dz_L}{z_L^{d+1-3\nu}}K_\nu(k_1 z_L)K_\nu(k_2 z_L)J_\nu(p z_L)
%\\
\propto&\;k_2^{(d+2)/2} p^{(d-2)/2}
{}_2F_1\left(-1,\frac{d-4}{2};\frac{d}{2};-\frac{p^2}{k_2^2}\right)
\ee 
where an appropriate regularization is assumed to have been applied in the computation.\footnote{
The proportionality factor in \eqref{eq: soft limit of transition amplitude} is
\be 
-i\sqrt{2}[-\bq\.\e^{h_1}(\bk_1)][\e^{h^*}(-\bq)\.\e^{h_2}(\bk_2)]\frac{\Gamma \left( \frac{d-2}{2}\right)\Gamma \left(-1\right) \Gamma \left(\frac{d-4}{2}\right)}{2^{(12-d)/2 } \Gamma (d/2)}
\ee 
which is clearly infinite, as evident from the $\G(-1)$ term. This is of course expected, as the limit of the transition amplitude is actually divergent and hence the result to be used is regularization-dependent. Here, we simply left divergent piece as it is; in practice, one could simply use a dimensional regularization which would merely shift the gamma function away from the pole.}

By inserting this into the former formula, we obtain the soft limit in the following schematic form:
\begin{multline}
\begin{aligned}
	{}&\text{Soft limit of}\\{}&\text{4-pnt diagram}
\end{aligned}
\;=\;
\text{\scalebox{1.5}{$\int$}}\begin{aligned}
	{}&\text{3-pnt transition}\\{}&\text{amplitude}
\end{aligned}\;\x\;{}_2F_1
\\+\;
\begin{aligned}
	{}&\text{Intrinsically}\\{}&\text{AdS contributions}
\end{aligned}
\end{multline}

\section{Generalization to Higher Point Amplitudes}
\label{sec:higher_point} 
In the last section, we discussed and demonstrated how different approaches to the computation of the soft limit of four point AdS amplitudes can lead to different looking results. Indeed, the differential representation which repackages different contributions into an operator acting on a single integral leads to the most compact form, whereas the direct approach that relies on the brute force computation of standard Feynman propagators yields the result that is arguably the simplest to work with, especially for numerical computations. However, despite being far more complicated in the way the terms are presented, the unitary decomposition approach that we advocate has the conceptual advantage of separating the degrees of freedom inherent to the curved space (i.e. longitudinal polarizations) from the rest. In addition, we will see in this section that unitary decomposition scales well with the higher point amplitudes, and in fact becomes a viable option even for computation purposes of the soft limit of higher point gluon amplitudes.

Let us now consider an $n+1$ point Witten diagram in which a soft leg of momentum $k_0$ and helicity $h_0$ enters through a 3-pt vertex. We can illustrate diagrammatically that we have the following decomposition
%\vspace*{-2em}
\be 
\label{eq: decomposition}
\hspace*{-.1in}
\tikzimage{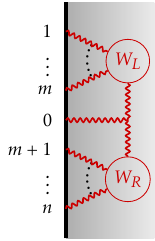}\quad\text{\scalebox{1.5}{$=$}}\hspace*{-1em}&\tikzimage{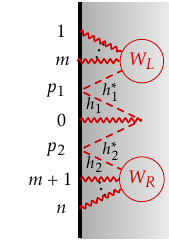}\quad\text{\scalebox{1.5}{$+$}}\hspace*{-1em}\tikzimage{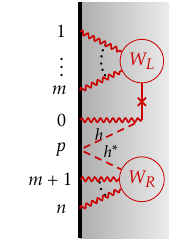}
\\
\text{\scalebox{1.5}{$+$}}&\tikzimage{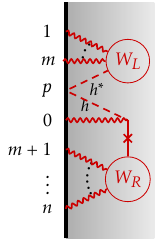}\quad\text{\scalebox{1.5}{$+$}}\tikzimage{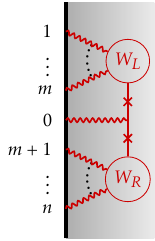}
\ee 
If we take the $\bk_0\rightarrow 0$ limit (hence take $0^{\text{th}}$ leg soft), we compute that the last piece is subleading by two orders hence we can drop it at the leading order. Furthermore, if we now introduce a \emph{boundary-to-boundary connector}\footnote{For compactness, we define $	p_>\coloneqq\max(p_1,p_2)$ and $p_<\coloneqq\min(p_1,p_2)$.}

\be
\cC(p_1,p_2)\coloneqq\begin{aligned}
	\frac{2^{2-d} \Gamma \left(\frac{3}{2}-\frac{d}{4}\right) \Gamma \left(\frac{d}{2}-1\right) p_>^{d-4} p_<^{d/2-1} }{\Gamma \left(\frac{3(d-2)}{4}\right)}
	\\\x\,_2\tilde{F}_1\left(\frac{6-d}{4},\frac{5}{2}-\frac{3 d}{4};\frac{d}{2};\frac{p_<^2}{p_>^2}\right)
\end{aligned}\doteq\tikzimage{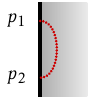}
\ee 
and a \emph{boundary-to-bulk connector}\footnote{We refrain from using the term propagator for these objects as they are merely integration kernels. We also note that bulk to boundary connectors are inherently curved space contributions that vanish in flat space limit.}
\be 
&\begin{aligned}
	\cC_i(\bk,p,z)\coloneqq{}&{}\bk_i z^{d/2-1}\frac{2^{2-\frac{3 d}{2}} p^{-5+(3d/2)} \Gamma \left(\frac{5}{2}-\frac{3 d}{4}\right) \Gamma \left(\frac{d-2}{2}\right)}{(d-2)\Gamma \left(\frac{5
		(d-2)}{4}\right)}\\{}&+\bk_i z^{4-d}\bigg[
\frac{ \, _1F_2\left(\frac{5}{2}-\frac{3 d}{4};\frac{7}{2}-\frac{3 d}{4},\frac{d}{2};-\frac{1}{4} p^2 z^2\right)}{(d-2)^2(3
	d-10)}\\{}&-\frac{ \, _1F_2\left(\frac{3}{2}-\frac{d}{4};\frac{5}{2}-\frac{d}{4},\frac{d}{2};-\frac{1}{4} p^2
	z^2\right)}{(d-2)^2(d-6)}
\bigg]
\end{aligned}
\\
&\phantom{\hspace*{4.2em}}\doteq\tikzimage{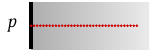}
\ee 
we arrive at the following form for the initial expression by carrying out the bulk-point integration at leading order in $\bk_0$ in the first three diagrams:\footnote{
Here, we define $c_1=-i2\sqrt{2}\bk_L\.\e^{h_0}(\bk_0)$, $c_2^h=\frac{i}{\sqrt{2}}\e^h(\bk_R)\.\e^{h_0}(\bk_0)$, and $c_3^h=-\frac{i}{\sqrt{2}}\e^{h}(\bk_L)\.\e^{h_0}(\bk_0)$ for brevity. The sums $\bk_L=-\sum\limits_{i=1}^m\bk_i$  and $\bk_R=-\sum\limits_{i=m+1}^n\bk_i$ are the momentum of the normalizable mode (or connector) that gets attached to $W_L$ and $W_R$ respectively.
}
\begin{multline}
\label{eq:diagrammatic form of soft limit}
\tikzimage{WittenGenericOne}\quad\text{\scalebox{1.5}{$=$}}\quad c_1\hspace*{-1em}\tikzimage{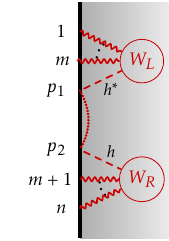}
\\\text{\scalebox{1.5}{$+$}}\;
c_2^h\hspace*{-1em}\tikzimage{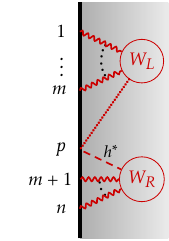}\;\text{\scalebox{1.5}{$+$}}\;c_3^{h^*}\hspace*{-.5em}\tikzimage{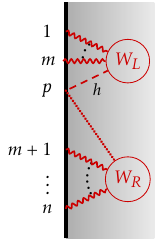}
\;\text{\scalebox{1.5}{$+$}}\;\cO\left(k_0\right)
\end{multline} 
Although we can write down the equations directly,\footnote{For instance, the first term on the right hand side of \equref{eq:diagrammatic form of soft limit} reads as
\begin{multline}
	\lim\limits_{\bk_0\rightarrow0}W(\bk_0^{h_0},\bk_1^{h_1},\dots,\bk_n^{h_h})\,\supset\, -i2\sqrt{2}\bk_L\.\e^{h_0}(\bk_0)k_0^{(d-2)/2}\sum\limits_{h}\\\x\int\frac{p_Lp_Rdp_Ldp_R}{4\left(k_L^2+p_L^2+i\e\right)\left(k_R^2+p_R^2+i\e\right)}\cC(p_L,p_R)
	\\\x
	W(\bk_L^{h^*},p_L;\bk_1^{h_1},\dots,\bk_m^{h_m})
	\\\x W(\bk_R^*,p_R;\bk_{m+1}^{h_{m+1}},\dots,\bk_n^{h_n})
\end{multline}
} we will rather focus on the diagrammatic notation as it will help us reveal the structural behavior in the soft limit in a clearer way. Namely, diagrammatically considering the soft limit $\bk_0\to0$ of other Witten diagrams in the same $(n+1)$-\emph{amplitude} (in a similar fashion to \equref{eq:diagrammatic form of soft limit}) and summing them seem to suggest the following observations:
\begin{enumerate}
	\item The soft limit of $n+1$ point diagram contains a piece (first term in right hand side) which is simply integration of lower point diagrams (with a leg made off-shell) against a kernel. In fact, since we sum over all diagrams in the $n+1$ point amplitude, we might expect that these terms add up to the integration of products of lower-point transition amplitudes agains a kernel.
	\item The resemblance of the above-mentioned term with unitary-cuts\footnote{In flat space, the unitarity cuts (i.e. Cutkosky rules) naturally arise because the decomposition of the Feynman propagator into Weightman functions in any given diagram basically rewrites it as a sum over products of lower point diagrams, analogous to our expressions above. The generalization of unitarity cuts into curved spaces has been investigated in the context of (A)dS, again with a remarkable yet expected similarity to our decomposition here (albeit in the correlator language); for further details, we refer the reader to \cite{Meltzer:2020qbr}.} is natural: by using the unitary decomposition of the propagator, we managed to relate higher point amplitudes to products of lower point ones, in an analogous fashion to the Cutkosky cuts.
	\item The inherently AdS pieces in the soft limit of $n+1$ point diagram can not be rewritten as products of lower point amplitudes, at least in the traditional sense since some of the bulk-to-boundary propagators are replaced by the bulk-to-boundary connectors.\footnote{We leave an in-depth analysis of these connectors to a future work.} 
	\item The suppression of different terms in the soft limit are of the same severity, i.e. all of them go as $(\bk_0)^{0}$. This is consistent with our main point in the introduction, i.e. unlike the flat space where particular diagrams are dominant in the soft limit (making Weinberg's soft theorem conceptually easier to grasp), and a similar analogy is missing in AdS. Conceptually, it is straightforward to understand this as all denominators in all Witten diagrams are nonsingular in soft limit, hence the amplitudes should scale $\cO(k^0)$ as $\bk\to 0$. 
	\item The analysis of \equref{eq:diagrammatic form of soft limit} gets corrections as there are also \emph{the edge cases}, where the soft leg is not connected to two bulk-to-bulk propagators; diagrammatically,
	\be 
	\label{eq: decomposition edge}
	\hspace*{-.1in}
	\tikzimage{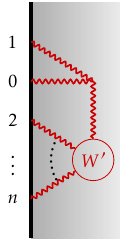}=&\tikzimage{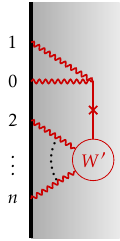}+\tikzimage{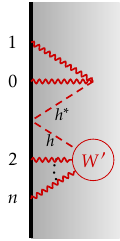}
	\ee 
	These diagrams do bring \emph{corrections} to our claims in the first items; for instance, summation of all diagrams (including the edge cases) is not warranted to yield exactly-and-only the integration of lower point transition amplitudes. Nevertheless, the ratio of edge cases in the total number of diagrams is suppressed in $1/n$, hence we believe that our claims are relevant for sufficiently high point amplitudes. Schematically, we \emph{conjecture} that
\begin{center}
	\scalebox{.85}{$
		\begin{aligned}
			\begin{aligned}
				{}&\text{Soft limit of}\\{}&\text{(n+1) point}\\{}&\text{diagram}
			\end{aligned}\quad\xrightarrow{\phantom{aaaaaaaa}}\quad\text{\scalebox{1.6}{$\sum\limits_m$}}\;
			\text{\scalebox{2.6}{$\int$}}\begin{aligned}
				{}&\text{m point}\\{}&\text{transition}\\{}&\text{amplitude}
			\end{aligned}\;\x\;\begin{aligned}
				{}&\text{n-m+1 point}\\{}&\text{transition}\\{}&\text{amplitude}
			\end{aligned}
			\\\\
			\text{\scalebox{1.6}{$+$}}\quad\text{curved-space contributions}
			\\\\
			\text{\scalebox{1.6}{$+$}}\quad\cO\left(\frac{1}{n}\right)\text{ corrections}
		\end{aligned}
		$}
\end{center}
which should hold in all dimensions.
\end{enumerate}

\section{Discussions}
\label{sec:conclusion}
In this work, we explore the soft limit behavior of gluon amplitudes in (Anti)-de Sitter space, focusing on the ways these amplitudes factorize. Unlike their flat counterparts, where soft limits yield neatly factorizable structures governed by unitarity,
curved space amplitudes do not factorize due to the additive nature of intrinsically (A)dS contributions. We show this explicitly by decomposing the AdS propagator into transverse and longitudinal modes, which allows us to systematically identify contributions specific to (A)dS, offering new insights into how gauge interactions are shaped in curved spacetime.

To address the non-trivial factorization properties of these amplitudes, we survey different representations of the propagator and employ unitary decomposition techniques. This approach reveals that AdS soft behavior diverges from familiar flat-space soft theorems, resulting instead in expressions linking lower-point transition amplitudes with curved-space corrections. 

There are several promising avenues for extending our current work.  One such direction is the use of soft limits in conjunction with on-shell recursion relations. These techniques have proven particularly effective in flat-space theories involving spin, with some initial successes observed in curved spacetime as well \cite{Raju:2011mp, Raju:2012zr, Albayrak:2023jzl}. Exploring soft recursion within this curved context could reveal insights analogous to those found in flat space \cite{Cheung:2015ota}. In a related vein, work has already been done in scalar theories that combine soft theorems with singularity structures using recursion relations \cite{Bittermann:2022nfh}. Additionally, constructing inverse soft limits—drawing inspiration from BCFW recursion—could potentially pave the way for developing new approaches to gauge-theory and gravity amplitudes \cite{Boucher-Veronneau:2011rwd}.

Another direction to extend our exploration is the generalization of soft limits to loop level. Practically, understanding soft limits in (A)dS may lead to new computational efficiencies and assist in broader theoretical frameworks, especially in cosmology and quantum gravity. These avenues collectively enhance our understanding of gauge theory, holography, and field theory in non-flat geometries, expanding the toolkit for addressing fundamental questions in curved spacetime physics. We believe that our findings lay a foundation for further research in these areas.

\begin{acknowledgments}
S.K. thanks Chandramoli Chowdhury for collaboration during the early stage of this project. SK thanks Jinwei Chu, Austin Joyce, Sid Prabhu, and Xingkang Wang for conversation. S.A. was supported in the first part of this project by a VIDI grant of the Netherlands Organisation for Scientific Research (NWO) that is funded by the Dutch Ministry of Education, Culture and Science (OCW); and, in the second part, he is supported by the BAGEP Award of the Science Academy in Turkey as well as T\"UB{$\dot{\mathrm{I}}$}TAK (The Scientific and Technological Research Council of Turkey) 2232-b International Fellowship for Early Stage Researchers programme with project number 122c153. Furthermore, S.A. would like to thank the organizers of the Bootstrap 2023 at the Instituto Principia and the organizers of the Bootstrap 2024 at the Complutense University where parts of this work were completed. S.A. expresses gratitude to the Department of Physics at the University of Chicago for their generous hospitality during the course of this research. S.K. acknowledges the support and hospitality of the Amplitudes 2024 conference at the Institute for Advanced Study, as well as Nikhef, and Middle East Technical University, where part of this work was conducted.\end{acknowledgments}

\appendix 
\renewcommand{\thesection}{\Alph{section}}
\numberwithin{equation}{section}

\section{Wavefunction coefficients and correlators in (A)dS}
\label{wave}
In this section, we will briefly review wavefunction method and its relation to correlation functions in AdS and dS. We will work in the flat slices with the metrics
\be 
ds^2_\ds=L_\ds^2\frac{-d\eta^2+\eta_{ij}dx^i dx^j}{\eta^2}\,\quad 
ds^2_\ads=L_\ads^2\frac{dz^2+\eta_{ij}dx^i dx^j}{z^2}
\ee 
where $\eta$, $z$, and $\eta_{ij}$ are the conformal time, bulk radius, and the boundary metric respectively.\footnote{Even though the discussion can be generalized, we will stick to a Euclidean three dimensional boundary as it is the cosmologically relevant scenario.} For brevity we will consider an action of a single scalar field, though one can generalize to arbitrarily many (potentially spinning) fields.

Let us start with AdS partition function: we integrate over all field configurations which satisfy the chosen boundary conditions, i.e.
\be 
\label{eq: AdS partition function}
Z_\ads[\Phi(x)]=\int\limits_{\lim\limits_{z\rightarrow 0}\f(x,z)=z^{\#}\Phi(x)}\cD\f\; e^{-S_\ads[\f]}
\ee 
Clearly, $\Phi(x)$ is simply a Dirichlet boundary condition from the bulk perspective; from the boundary side, it actually corresponds to the coefficient of operator deformations of the CFT Lagrangian. With this perspective, one can compute the CFT correlation functions by taking functional derivatives of $Z_\ads$ with respect to these sources.\footnote{\label{footnote: AdS vs dS}
This AdS/CFT dictionary given in \cite{Witten:1998qj,Gubser:1998bc} yields the same boundary correlation functions obtained from the bulk ones via extrapolation \cite{Banks:1998dd}. As discussed in \cite{Harlow:2011ke}, this is no longer true in dS: we have two inequivalent ways to ``define'' boundary correlation functions from the bulk. We'll discuss this point below.
}

Unlike the $Z_\ads[\Phi(x)]=Z_\cft[\Phi(x)]$ equality of AdS/CFT dictionary, the dS/CFT correspondence states\footnote{
See \cite{Witten:2001kn,Strominger:2001pn} for the early conjectures of dS/CFT dictionary. The equality between the CFT partition function and dS wavefunction in \equref{eq: dS/CFT relation} is proposed in \cite{Maldacena:2002vr}.}
\be 
\label{eq: dS/CFT relation}
\Psi_\ds[\Phi(x)]=Z_\cft[\Phi(x)]
\ee 
One might attribute the difference between AdS/CFT and dS/CFT dictionaries to the different signatures of the boundary (or the absence of time direction in the AdS if both boundaries are Euclidean).\footnote{
One can write down the analogs of the standard wavefunction defined on a constant time spacelike hypersurface in other quantizations (such as radial quantization) with other kinds of hypersurfaces (such as timelike in case of a Lorentzian AdS): see \cite{Hartle:1983ai} for the related Hartle-Hawking wave function, and \cite{Witten:1998wy,Lifschytz:2000bj,deBoer:1999tgo} for more on this topic. Thus the signature of the hypersurface or the kind of the quantization is not the main issue here.} However, the real culprit is the role of the boundary fields: in the AdS, they are literally static boundary conditions from the bulk perspective, whereas in dS they are the future values of the dynamic bulk fields and hence are allowed to fluctuate.\footnote{To our knowledge, this is first pointed out in \cite{Maldacena:2002vr}, systematically  analyzed in \cite{Harlow:2011ke}, and is used to address the subtleties of dS Ward identities in \cite{Ghosh:2014kba}.} Indeed, one has to \emph{integrate over} these dynamic values to compute the correlation functions at the boundary of dS:
\be 
\<\Phi(x_1)\dots\Phi(x_n)\>_\ds^{(1)}=\int \cD\Phi\; \Phi(x_1)\dots\Phi(x_n)\; \abs{\Psi_\ds[\Phi(x)]}^2
\ee 
which is precisely the Born rule from the boundary perspective. Indeed, at any constant-time $\eta=\eta_0$ slice of dS, the correlation functions are given by this rule hence the boundary correlator above can be seen as the boundary extrapolation of the bulk correlation function, schematically
\be 
\<\Phi(x_1)\dots\Phi(x_n)\>_\ds^{(1)}=\lim\limits_{\eta_0\rightarrow 0}\eta_0^{\#}\<\phi(x_1,\eta_0)\dots\phi(x_n,\eta_0)\>_\ds
\ee 
However, not unlike the AdS case, we can also compute the boundary correlators from the functional derivatives of the boundary partition function
\be 
\<\Phi(x_1)\dots\Phi(x_n)\>_{\ds}^{(2)}\sim&\left[\frac{\de}{\de \Phi(x_1)}\cdots \frac{\de}{\de \Phi(x_n)}\right]Z_\cft
\\
\sim&\left[\frac{\de}{\de \Phi(x_1)}\cdots \frac{\de}{\de \Phi(x_n)}\right]\Psi_{\ds}
\ee 
Clearly, these two definitions (bulk extrapolation vs boundary derivatives) do not match in dS.\footnote{They do match in AdS, see footnote~\ref{footnote: AdS vs dS}.} On the other hand, if one were to start with the boundary correlators in AdS and do the analytic continuation\footnote{Besides taking $ds^2_\ads$ to $ds^2_\ds$, this continuation also ensures that AdS fields obeying bulk regularity as $z\rightarrow\infty$ analytically continue to dS fields obeying Bunch Davis boundary conditions as $\eta\rightarrow-\infty$; see \cite{Ghosh:2014kba} for more details.}
\be 
L_\ds=-i L_\ads\;,\quad \eta=i z\;,
\ee 
one ends up with the boundary correlators of the second method; schematically,
\be 
\left[\frac{\de}{\de \Phi(x_1)}\cdots \frac{\de}{\de \Phi(x_n)}\right]\Psi_{\ds}\;\sim\;& \<\Phi(x_1)\dots\Phi(x_n)\>_{\ds}^{(2)}
\\
&\Bigg\downarrow\text{\small (analytic  continuation)}
\\\\[-.1in]
\;\sim\;&\<\Phi(x_1)\dots\Phi(x_n)\>_{\ads}
\\\\[-.1in]
\;\sim\;& \left[\frac{\de}{\de \Phi(x_1)}\cdots \frac{\de}{\de \Phi(x_n)}\right]Z_{\ads}
\ee 
Therefore, one usually says that the AdS partition function analytically continues to the dS wavefunction.\footnote{
Such statements are dependent on the scheme one works with. For instance, if we choose to work with the holographic renormalization formalism of \cite{Heemskerk:2010hk}, we can decompose the AdS partition function in \equref{eq: AdS partition function} into three parts with respect to a flat slicing at $z=l$:
\be 
Z_\ads[\Phi(x)]=&\int\cD\tl\f \left[\;\int\limits_{\lim\limits_{z\rightarrow 0}\f(x,z)=z^{\#}\Phi(x)}^{\f(x,l)=l^{\#}\tl\f(x)}\cD\f\evaluated_{z<l}\; e^{-S_\ads[\f]\evaluated_{z<l}}\right]
\\&\x 
\left[\;\int\limits_{\f(x,l)=l^{\#}\tl\f(x)}\cD\f\evaluated_{z>l}\;
e^{-S_\ads[\f]\evaluated_{z>l}}\right]
\\
=&\int\cD\tl\Phi \Psi_{\text{\tiny UV}}[\tl\f,l]
\ee 
}

\section{On perturbative unitarity and the decomposition of the propagator}
\label{appendix unitarity}
Consider the Feynman propagator of a spin-1 field in flat space:\footnote{We are suppressing any color factors.}
 \be 
 \label{eq:flat space propagator}
 \Pi_1^{\mu\nu}(p)=\frac{g^{\mu\nu}-(1-\xi)p^\mu p^\nu /p^2}{p^2-m^2+i\e}
 \ee 
 For a massive field, $\xi=0$ and the numerator is a sum over physical spin states. For a massless field, $\xi$ is an arbitrary parameter but gauge invariance guarantees that the final result is independent of the value of $\xi$. In fact, similar to the massive case, we can write the numerator as a sum over physical states. This is no coincidence as unitarity is actually the certificate of the equality
 \be 
 \label{eq: physical form of flat space propagator}
 \Pi_{\mu\nu}(p)=\frac{\sum\limits_{h}\e^h_\mu(p)\left[\e^h_{\nu}(p)\right]^*}{p^2-m^2+i\e}
 \ee 
 for any spin.\footnote{We recommend the nice discussion in \S24.1.3 of \cite{Schwartz:2014sze}.} Thus the Feynman propagator for a massless particle in flat space can be written as\footnote{Here we define $G^{\text{flat}}_{\mu\nu}(\bk_i,z,z')$ via the relation
\be 
G^{\text{flat}}_{\mu\nu}(x_\mu,x_\nu')\coloneqq\int\limits_{\R^{d-1,1}}\frac{d^dk}{(2\pi)^d} G^{\text{flat}}_{\mu\nu}(\bk,z,z')e^{i\bk\.(\bm{x}-\bm{x}')}
\ee  
where the last component of $\R^{d,1}$ vector $x_\mu$ is $z$, and the rest is $\bm{x}$.
}
\be 
\label{eq: flat space propagator}
G^{\text{flat}}_{\mu\nu}(\bk,z,z')=\sum\limits_h\int\limits_{\R}\frac{dp}{2\pi}
\frac{A^{h,\text{flat}}_\mu(\bk,p,z)\left[A^{h,\text{flat}}_\mu(\bk,p,z')\right]^*}{k^2+p^2+i\e}
\ee 
for the off-shell external states
\be 
\label{eq: flat space bulk to boundary}
A^{h,\text{flat}}_\mu(\bk,p,z)\coloneqq \e_\mu^h(\bk)e^{ipz}
\ee 
where we distinguish the last coordinate by labeling it as $z$ and the corresponding momentum as $p$. In summary, unitarity ensures that in flat space the Feynman propagator (i.e. bulk-to-bulk propagator) can be written as a sum over all possible helicity states of product of off-shell external states (i.e. bulk-to-boundary propagators) integrated against a measure that would put them on-shell via contour deformation.

An analog of this exists in AdS and is called the split-representation; indeed, we can write\footnote{For instance, see \cite{Meltzer:2019nbs} and the references therein for further details.}
\be 
G^{\text{AdS}}(\De,y,y')=\int\limits_{-\infty}^{\infty}\frac{\nu^2d\nu}{\pi}\int\limits_{\partial\text{AdS}} d^dx \frac{A_{\frac{d}{2}+i\nu}(x,y)A_{\frac{d}{2}+i\nu}(x,y')}{\nu^2+\left(\De-\frac{d}{2}\right)^2}
\ee 
where $G^{\text{AdS}}$ is the bulk-to-bulk propagator and $A$ is the bulk-to-boundary propagator for the bulk points $y$ and boundary points $x$. Indeed, $\nu$ integration can be seen as an analog of $p$ integration, as both encode the harmonic analysis of the respective groups.\footnote{In flat space, the isometry group is $\R^{d,1}$ and is parametrized by the momenta $\bk$: the Fourier transformation is literally a map between the group space and the representation space. In contrast, AdS isometry is the conformal group, hence the representations are paremetrized by the scaling dimension $\De$ and the spin. The relevant integration now is of the scaling dimension over the \emph{Euclidean principle series}.} However, the flip side of this story is that we do not expect to have an analogous form of \equref{eq: flat space propagator} in \emph{momentum space}, as it is not the respective representation space.

The situation is analogous in dS but there are further complications. In dS, the integration is over a manifold that does not respect the time reversal as we take $t_{\text{initial}}$ to $-\infty$ and $t_{\text{final}}$ to $0$ ($\infty$ in flat space); in other words, the analogs of Feynman's $i\epsilon$ prescriptions lead to regulated evolution operators that are no longer unitary.\footnote{See \cite{Albayrak:2023hie} for a discussion of these points in detail.} There are alternative prescriptions that yield a regularized unitary evolution operators, but there is no canonical unique choice and the restrictions due to causality on this space of choices are not fully understood. In short, we do not expect an analog of the statement at the end of the paragraph after \equref{eq: flat space bulk to boundary}, neither in AdS nor in dS.

Despite all these differences, there are still advantages to bring the (A)dS propagator to a form similar to the flat space one in \equref{eq: flat space propagator}, as we did in \equref{eq: decomposition of AdS propagator}. Firstly, this allows us to keep working with boundary momentum as variables, directly connectable to cosmological and other experimental data. Secondly, this makes the connection with the flat space results easier to analyze, as pieces that vanish in the flat space limit are now manifestly separated. Lastly, although we are not doing it in this paper, it is possible to turn the bulk-to-bulk propagator on-shell by focusing on the pole of the denominator,\footnote{This is effectively switching to old-fashioned perturbation theory by making sure all internal legs are on-shell, similar to the external ones.} and using the identity
\be 
z^\nu J_\nu(-ik z)=\frac{ie^{-i\pi\nu/2}}{\pi}z^\nu K_\nu\left(-kz\right)-\frac{ie^{-3i\pi\nu/2}}{\pi}z^\nu K_\nu\left(kz\right)
\ee 
that is valid for the relevant $z>0$ range. By this, we can decompose the propagator to ``negative frequency'' and ``positive frequency'' bulk-to-boundary propagators, and carry out the (A)dS analysis in terms of propagation of such modes.\footnote{Although these modes are basically radial momenta propagation in AdS, they are literally positive and negative energy modes in dS, hence are more analogous to the flat space analysis.} 
\section{On the soft limit of Bessel type integrals}
\label{appendix bessels}
In the analysis of AdS$_{d+1}$ amplitudes in momentum space for arbitrary $d$, it is convenient to define the following bulk integrals
\bea[eq: bessel integrals]
\jjk_{\l,\mu}(a,b,c)\coloneqq& \int\limits_0^\infty dz z^{\lambda-1}J_\mu(a z)J_\mu(b z)K_\mu(c z)
\\
\jkk_{\l,\mu}(a,b,c)\coloneqq& \int\limits_0^\infty dz z^{\lambda-1}J_\mu(a z)K_\mu(b z)K_\mu(c z)
\\
\jk^{(1)}_{\l,\mu}(a,b,l)\coloneqq& \int\limits_{0}^{l}dz
z^{\l-1}
J_\mu (a z)
K_\mu (b z)
\\
\jk^{(2)}_{\l,\mu}(a,b,l)\coloneqq& \int\limits_{l}^{\infty}dz
z^{\l-1}
J_\mu (a z)
K_\mu (b z)
\\
\kk^{(1)}_{\l,\mu}(a,b,l)\coloneqq& \int\limits_{0}^{l}dz
z^{\l-1}
K_\mu (a z)
K_\mu (b z)
\\
\kk^{(2)}_{\l,\mu}(a,b,l)\coloneqq& \int\limits_{l}^{\infty}dz
z^{\l-1}
K_\mu (a z)
K_\mu (b z)
\eea 
which were extensively used in the body of this paper. In this appendix, we analyze these expressions and their limiting behavior.

First two integrals above can be expressed in terms of Appell’s hypergeometric function as\footnote{See Appendix of \cite{Albayrak:2020bso} for further details.}
\begin{multline}
	\label{eq: JJK in arbitrary dimensions}
	\jjk_{\l,\mu}(a,b,c)=\frac{2^{\lambda-2}a^{\mu } b^{\mu }  \Gamma \left(\frac{\lambda +3\mu}{2}\right )\Gamma \left(\frac{\lambda +\mu}{2}\right )}{c^{ \lambda + 2\mu}\Gamma (\mu +1)^2}
	\\\x F_4\left(\frac{\lambda +\mu}{2}
	,\frac{\lambda +3\mu}{2};\mu +1,\mu +1;-\frac{a^2}{c^2},-\frac{b^2}{c^2}\right)
\end{multline} 
and
\begin{multline}
	\jkk_{\l,\mu}(a,b,c)=\Bigg[\frac{\Gamma (\mu )\Gamma \left(\frac{\lambda -\mu		}{2}\right) \Gamma \left(\frac{\lambda +\mu }{2}\right)}{2^{3-\lambda } c^{\lambda } \left(\frac{c}{a}\right)^{\mu }
		\left(\frac{b}{c}\right)^{\mu }\Gamma (\mu +1)}
	\\\x F_4\left(\frac{\lambda -\mu}{2}
	,\frac{\lambda +\mu}{2};1+\mu ,1-\mu;-\frac{a^2}{c^2},\frac{b^2}{c^2}\right)\Bigg]
	\\+
	\Bigg[\frac{\Gamma (-\mu )\Gamma \left(\frac{\lambda +\mu}{2}\right) \Gamma \left(\frac{\lambda +3\mu}{2}\right)}{2^{3-\lambda } c^{\lambda } \left(\frac{c}{a}\right)^{\mu }
		\left(\frac{b}{c}\right)^{-\mu }\Gamma (\mu +1)}
	\\\x F_4\left(\frac{\lambda +\mu}{2}
	,\frac{\lambda +3\mu}{2};1+\mu ,1+\mu;-\frac{a^2}{c^2},\frac{b^2}{c^2}\right)\Bigg]
\end{multline}
To consider the soft limit of $\jkk$, we use the fact that
\be
\label{eq: transition to 2F1} 
F_4\left(a,b;c,d;x,0\right)={}_2F_1\left(a,b,c,x\right)
\ee 
hence, for $b,\mu>0$, we obtain
\begin{multline}
	\lim\limits_{b\rightarrow 0}\jkk_{\l,\mu}(a,b,c)=b^{-\mu}\bigg[\frac{a^\mu\Gamma (\mu )\Gamma \left(\frac{\lambda -\mu		}{2}\right) \Gamma \left(\frac{\lambda +\mu }{2}\right)}{2^{3-\lambda } c^{\lambda } \Gamma (\mu +1)}
	\\\x {}_2F_1\left(\frac{\lambda -\mu}{2}
	,\frac{\lambda +\mu}{2};1+\mu;-\frac{a^2}{c^2}\right)+\cO(b)\bigg]
\end{multline}
In comparison, to obtain the soft limit of $\jjk$, we first use the relation\footnote{See eqn.9 of \cite{bateman1953higher} in page 240. Note that there is a minus sign error in that equation (and the erratum does not refer to it); the correct version of the equation can be found in the original book of Appell (available in French only), see eqn. 37 in page 26 of \cite{appell1926fonctions}.}
\begin{multline}
	F_4(a,b,c,d;x,y)=\frac{\G(d)\G(b-a)}{\G(d-a)\G(b)}(-y)^{-a}\\\x F_4(a,a+1-d,c,a+1-b;x/y,1/y)\\
	+\frac{\G(d)\G(a-b)}{\G(d-b)\G(a)}(-y)^{-b}\\\x F_4(b+1-d,b,c,b+1-a;x/y,1/y)
\end{multline}
which leads to 
\begin{multline}	
\jjk_{\l,\mu}(a,b,c)=\bigg[\frac{2^{\lambda-2}a^{\mu } }{c^{\mu}b^{\l}}
	\frac{\G(\frac{\l+\mu}{2})}{\mu \G(\frac{2-\lambda +\mu}{2})}\\\x F_4(\frac{\lambda +\mu}{2},\frac{\lambda -\mu}{2},\mu +1,1-\mu;\frac{a^2}{b^2},-\frac{c^2}{b^2})
	\\+\frac{2^{\lambda-2}a^{\mu }c^{\mu}}{b^{\l+2\mu }}\frac{\G(\frac{\l+3\mu}{2})\G(-\mu)}{\G(\frac{2-\lambda -\mu}{2})\G(1+\mu)}\\\x F_4(\frac{\lambda +\mu}{2},\frac{\lambda +3\mu}{2},\mu +1,1+\mu;\frac{a^2}{b^2},-\frac{c^2}{b^2})
	\bigg]
\end{multline} 
In the soft limit for $c,\mu>0$, we then see that
\begin{multline}
	\lim\limits_{c\rightarrowtail0}\jjk_{\l,\mu}(a,b,c)=c^{-\mu}\bigg[
	\frac{2^{\lambda-2}a^{\mu } \G(\frac{\l+\mu}{2})}{b^{\l}\mu \G(\frac{2-\lambda +\mu}{2})}\\\x{}_2F_1(\frac{\lambda +\mu}{2},\frac{\lambda -\mu}{2},\mu +1;\frac{a^2}{b^2})+\cO(c)
	\bigg]
\end{multline}
where we also used \equref{eq: transition to 2F1}.

We note the following useful identities\footnote{
	We warn the reader that Mathematica is unable to show these identities analytically, though it does verify them numerically. One can derive those by starting with eqn. 16 in page 124 of \cite{bateman1953higher}.
}
\bea 
		\, _2F_1\left(\frac{\lambda -\mu }{2},\frac{\lambda +\mu }{2};\mu +1;-\frac{b}{a}\right)=\G(1+\mu)\left(\frac{a}{a+b}\right)^{\lambda /2}\nn\\\x
		\left(\frac{b+a}{b}\right)^{\mu /2} P_{\frac{\mu -\lambda }{2}}^{-\mu }\left(\frac{a-b}{a+b}\right)\quad\text{ for }a,b>0\text{ and }\lambda,\mu\in\R
\\
		\, _2F_1\left(\frac{\lambda -\mu }{2},\frac{\lambda +\mu }{2};\mu +1;\frac{b}{a}\right)=\Gamma (\mu +1) \left(\frac{a}{a-b}\right)^{\lambda /2}
		\nn\\\x
		\left(\frac{b-a}{b}\right)^{\mu /2} P_{\frac{\mu -\lambda }{2}}^{-\mu
		}\left(\frac{a+b}{a-b}\right)\quad\text{ for }a>b>0\text{ and }\lambda,\mu\in\R
\eea
with which we get our final expressions
\bea[eq: soft limit of JKK and JJK]
\lim\limits_{q\rightarrow 0}\jkk_{\l,\mu}(p_J,p_K,q)=q^{-\mu}\bigg[\frac{\Gamma \left(\frac{\lambda -\mu		}{2}\right) \Gamma \left(\frac{\lambda +\mu }{2}\right)\Gamma (\mu )}{2^{3-\lambda }}
\nn\\\x\left(p_J^2+p_K^2\right)^{(\mu-\lambda) /2} P_{\frac{\mu -\lambda }{2}}^{-\mu }\left(\frac{p_K^2-p_J^2}{p_K^2+p_J^2}\right)+\cO(q)\bigg]
\\\nn\\
\lim\limits_{q\rightarrow 0}\jjk_{\l,\mu}(p_1,p_2,q)=q^{-\mu}\bigg[
\frac{(-1)^{\mu/2}\G(\frac{\l+\mu}{2})\Gamma (\mu)}{2^{2-\lambda}\G(\frac{2-\lambda +\mu}{2})}\nn\\\x\left(P_{12}^2-p_{12}^2\right)^{(\mu-\lambda) /2} P_{\frac{\mu -\lambda }{2}}^{-\mu
}\left(\frac{P_{12}^2+p_{12}^2}{P_{12}^2-p_{12}^2}\right)+\cO(q)
\bigg]
\eea 
for $p_{12}=\min(p_1,p_2)$ and $P_{12}=\max(p_1,p_2)$, and where $P^a_b(x)$ is the associated Legendre function of the first kind.

The last four integrals in \equref{eq: bessel integrals} are harder to compute. We know that their sum are proportional to a hypergeometric function, i.e.\footnote{These follow from \emph{the Weber-Schafheitlin integral} \cite{https://doi.org/10.1112/plms/s2-40.1.37}
	\begin{multline}
		\int\limits_0^\infty t^{\l-1}K_\mu(at)J_{\nu}(bt)dt=\frac{b^\nu\G\left(\frac{\l+\mu+\nu}{2}\right)\G\left(\frac{\l-\mu+\nu}{2}\right)}{2^{2-\l}a^{\nu+\l}\G(\nu+1)}\\\x F\left(\frac{\l+\nu+\mu}{2},\frac{\l-\mu+\nu}{2};\nu+1;-\frac{b^2}{a^2}\right)
	\end{multline}
and the \emph{Sonine's integral} \cite{zwillinger2014table}
\begin{multline}
\int\limits_0^\infty t^{\l-1}K_\mu(at)K_{\nu}(bt)dt=
\frac{b^\nu\G\left(\frac{\l+\mu+\nu}{2}\right)\G\left(\frac{\l-\mu+\nu}{2}\right)\G\left(\frac{\l+\mu-\nu}{2}\right)\G\left(\frac{\l-\mu-\nu}{2}\right)}{2^{3-\l}a^{\nu+\l}\G(\l)}\\\x F\left(\frac{\l+\mu+\nu}{2},\frac{\l-\mu+\nu}{2};\l;1-\frac{b^2}{a^2}\right)
\end{multline}
}
\bea 
{}&\jk^{(1)}_{\l,\mu}(a,b,l)+\jk^{(2)}_{\l,\mu}(a,b,l)=\frac{a^\mu\G\left(\mu+\frac{\l}{2}\right)}{2^{2-\l}b^{\mu+\l}\G(\mu+1)}\nn\\{}&\qquad\x\G\left(\frac{\l}{2}\right) F\left(\mu+\frac{\l}{2},\frac{\l}{2};\mu+1;-\frac{a^2}{b^2}\right)
\\{}&
\kk^{(1)}_{\l,\mu}(a,b,l)+\kk^{(2)}_{\l,\mu}(a,b,l)=
\frac{b^\mu\G\left(\mu+\frac{\l}{2}\right)\G\left(-\mu+\frac{\l}{2}\right)}{2^{3-\l}a^{\mu+\l}\G(\l)}\nn\\{}&\qquad\x\left(\G\left(\frac{\l}{2}\right)\right)^2 F\left(\mu+\frac{\l}{2},\frac{\l}{2};\l;1-\frac{b^2}{a^2}\right)\\\nn
\eea 
however, a general method to compute individual integrals for arbitrary symbolic arguments is unbeknownst to us. Therefore, we will leave them as they are: note that they actually become computable once one focuses on a specific set of $\{\l,\mu\}$.\footnote{For instance,
	\begin{multline}
		\jk^{(1)}_{1,1/2}(a,b,l)=-\frac{i}{2 \sqrt{a	b}}\Big(-\text{Chi}(-i a-b)+\text{Chi}(i a-b)+\text{Shi}(i
		a-b)\\+\text{Shi}(i a+b)+\log (-b-i a)-\log (-b+i a)\Big)
	\end{multline}
	for sinh and cosh integrals \emph{Shi} and \emph{Chi}.
} In comparison, their soft limit can be computed in closed form for arbitrary $\{\l,\mu\}$; to see this, we simply expand the integrand in soft limit and use relevant general formula\footnote{Note the relations
\bea 
\int\limits_{0}^l z^{k-\mu-1}J_\mu(a z)dz={}&{}\frac{2^{-m} a^m l^k \,
	_1F_2\left(\frac{k}{2};\frac{k}{2}+1,m+1;-\frac{1}{4} a^2
	l^2\right)}{k \Gamma (m+1)}\\
\int\limits_{l}^\infty z^{k+\half}J_\mu(a z)dz={}&{}\frac{2^{k+\frac{1}{2}} a^{-k-\frac{3}{2}} \Gamma \left(\frac{1}{4} (2 k+2 m+3)\right)}{\Gamma \left(\frac{1}{4} (-2 k+2 m+1)\right)}\nn\\{}&{}\hspace*{-7em}-\frac{2^{1-m} a^m l^{k+m+\frac{3}{2}} \,
	_1F_2\left(\frac{k}{2}+\frac{m}{2}+\frac{3}{4};\frac{k}{2}+\frac{m}{2}+\frac{7}{4},m+1;-\frac{1}{4} a^2 l^2\right)}{(2 k+2 m+3) \Gamma (m+1)}
\\
\int\limits_{0}^l z^{k-1}K_\mu(a z)dz={}&{}
\frac{\pi \Gamma\left(\frac{k-\mu }{2}\right)    \,_1\tilde{F}_2\left(\frac{k-\mu }{2};1-\mu ,\frac{k-\mu}{2}+1;\frac{1}{4}a^2 l^2\right)}{2^{2-\mu}l^{\mu-k}a^{\mu }\sin(\pi  \mu)}
\nn\\{}&{}+\;\mu\leftrightarrow -\mu
\\
\int\limits_{l}^\infty z^{k-1}K_\mu(a z)dz={}&{}
2^{k-2} a^{-k} \Gamma \left(\frac{k-\mu }{2}\right) \Gamma
\left(\frac{k+\mu }{2}\right)
\nn\\{}&{}\hspace*{-5em}+\left(\frac{\Gamma (\mu )\,
	_1F_2\left(\frac{k}{2}-\frac{\mu }{2};1-\mu
	,\frac{k}{2}-\frac{\mu }{2}+1;\frac{a^2 l^2}{4}\right)}{2^{1-\mu} a^{\mu } l^{\mu-k} (\mu
	-k)}+\;\mu\leftrightarrow -\mu\right)
\eea 
} to obtain
\begin{widetext}
\small
\bea[eq: limits of kk, jk, etc]
\lim\limits_{b\rightarrow0}\jk^{(1)}_{\l,\mu}(a,b,l)={}&{}\frac{a^{\mu } b^{-\mu } l^{\lambda } \Gamma (\mu ) \, _1F_2\left(\frac{\lambda }{2};\frac{\lambda }{2}+1,\mu +1;-\frac{1}{4} a^2 l^2\right)}{2
	\lambda  \Gamma (\mu +1)}
\\
\lim\limits_{b\rightarrow0}\jk^{(2)}_{\l,\mu}(a,b,l)={}&{}\frac{1}{4} a^{\mu } b^{-\mu } \Gamma (\mu ) \left(\frac{2^{\lambda } a^{-\lambda } \Gamma \left(\frac{\lambda }{2}\right)}{\Gamma
	\left(-\frac{\lambda }{2}+\mu +1\right)}-\frac{2 l^{\lambda } \, _1F_2\left(\frac{\lambda }{2};\frac{\lambda }{2}+1,\mu +1;-\frac{1}{4} a^2
	l^2\right)}{\lambda  \Gamma (\mu +1)}\right)
\\
\lim\limits_{b\rightarrow0}\kk^{(1)}_{\l,\mu}(a,b,l)={}&{}\frac{1}{4} l^{\lambda } \Gamma (\mu ) (a b)^{-\mu }
\left(\frac{4^{\mu } l^{-2 \mu } \Gamma (\mu ) }{\lambda -2 \mu
}\,
_1F_2\left(\frac{\lambda }{2}-\mu ;1-\mu ,\frac{\lambda
}{2}-\mu +1;\frac{a^2 l^2}{4}\right)+\frac{a^{2 \mu } \Gamma (-\mu ) \, _1F_2\left(\frac{\lambda
	}{2};\frac{\lambda }{2}+1,\mu +1;\frac{a^2
		l^2}{4}\right)}{\lambda }
\right)
\\
\lim\limits_{b\rightarrow0}\kk^{(2)}_{\l,\mu}(a,b,l)={}&{}\frac{\Gamma (\mu ) }{8(a b)^{\mu}}
\left(\frac{2^{2 \mu +1}
	\Gamma (\mu ) l^{\lambda -2 \mu } \, _1F_2\left(\frac{\lambda
	}{2}-\mu ;1-\mu ,\frac{\lambda }{2}-\mu +1;\frac{a^2
		l^2}{4}\right)}{2 \mu -\lambda }-\frac{2 a^{2 \mu } l^{\lambda
	} \Gamma (-\mu ) \, _1F_2\left(\frac{\lambda
	}{2};\frac{\lambda }{2}+1,\mu +1;\frac{a^2
		l^2}{4}\right)}{\lambda }+\frac{ \Gamma
		\left(\frac{\lambda }{2}\right) \Gamma
		\left(\frac{\lambda }{2}-\mu \right)}{2^{-\lambda }a^{\lambda-2 \mu}}\right)
\eea 
\normalsize
\end{widetext}

\begin{table*}
\caption{\label{table} Examples of three point amplitudes in various dimensions. Here $a,b,c$ are judiciously chosen kinematic variables respecting the permutation symmetry of the system: $a=\abs{\bk_1}+\abs{\bk_2}+\abs{\bk_3}$, $b=\abs{\bk_1} \abs{\bk_2} \abs{\bk_3}$, and $c=\abs{\bk_1} \abs{\bk_2} (\abs{\bk_1}+\abs{\bk_2})+\text{permutations}$}
	\begin{tabular}{ll}
\hline\hline\\[-1em] Gluon in AdS$_4$&		$\displaystyle\cA_3=\frac{1}{a}$\\\\[-1em]\hline\\[-1em] Gluon in AdS$_6$&$\displaystyle\cA_3=\frac{-a^3+4 b+c}{a^2}$\\\\[-1em]\hline\\[-1em] Graviton in AdS$_6$&$\displaystyle\cA_3=\frac{3 a^6-3 a^3 (8 b+3 c)+38 b^2+21 b c+3 c^2}{a^3}$\\\\[-1em]\hline\\[-1em] Graviton in AdS$_8$&$\displaystyle\cA_3=\frac{3 \left(-5 a^9+5 a^6 (12 b+5 c)-a^3 \left(226 b^2+165 b c+30
			c^2\right)+194 b^3+169 b^2 c+50 b c^2+5 c^3\right)}{a^4}$\\\\[-1em]\hline\hline
	\end{tabular}
\end{table*}

\section{Soft limit in special cases}
\label{sec:ads4three}

\subsection{AdS$_4$ computations}
The discussion in the body of the paper is purely agnostic to the spacetime dimension, making expressions rather technical and lengthy as a result. Once one sticks to a particular dimension, many of the formulae actually simplify drastically. We demonstrate this by focusing on $d=3$, which is a quite interesting case as it is the relevant dimension for (A)dS$_4$.

With $d=3$ (hence $\nu=1/2$), all Bessel functions simplify to exponentials and sinosoidal functions, with which the definitions of the bulk to boundary propagators take the simple form
\bea 
A_i^{h_a}(\bk_a,z)\coloneqq\e_i^{h_a}(\bk_a)\sqrt{\frac{\pi}{2}}e^{- k_a z}\doteq\tikzimage{kmode}
\\
\hat A_i^{h}(\bk,z)\coloneqq\e_i^{h}(\bk)\sqrt{\frac{2}{\pi }}\sin(k z)\doteq\tikzimage{jmode}\\
\hat A_i^{h}(\bk,p,z)\coloneqq\e_i^{h}(\bk)\sqrt{\frac{2k}{\pi p}}\sin(p z)\doteq\tikzimage{jmodeoff}
\eea 
where the bulk-to-bulk propagator in the unitary decomposition consequently becomes
\begin{multline}
	G_{ij}(\bk,z,z')=\frac{\bk_i\bk_j}{2 k^2}\sqrt{\frac{\min\left(z,z'\right)}{\max\left(z,z'\right)}}
	\\+\sum\limits_h \e_i^{h^*}(\bk) \e_j^{h}(\bk)\int\limits_{\R^+}\frac{dp}{\pi}\frac{\sin(pz)\sin(pz')}{k^2+p^2+i\e}\\
	\,\doteq\,\tikzimage{propagatorL}\,+\,\sum\limits_h\bigintssss\limits_{\R^+}\frac{pdp}{2k(k^2+p^2)}\,\tikzimage{propagatorFlat}
\end{multline}

One can now straightforwardly verify that many expressions above (especially those involving multiple complicated hypergeometric functions) become quite elementary; but more importantly, many essential integrals are now doable. For instance, even though we do not have an analytic formula for the definite version of Sonine's integral in general dimensions (as discussed in appendix~\ref{appendix bessels}), the relevant integrals are actually available in $d=3$, yielding
\bea 
\kk^{(1)}_{\l,\half}(a,b,l)=& \frac{\pi  (a+b)^{1-\lambda } (\Gamma (\lambda -1)-\Gamma
	(\lambda -1,(a+b) l))}{2 \sqrt{a b}}
\\
\kk^{(2)}_{\l,\half}(a,b,l)=& \frac{\pi  l^{\lambda -1} E_{2-\lambda }((a+b) l)}{2 \sqrt{a b}}
\eea 
where $E_n(x)$ is the \emph{exponential integral function}.

The availability of these integrals simplify various expressions; for instance, the s-channel Witten diagram for four gluons in \equref{eq: s channel witten diagram in unitary decomposition} becomes
\begin{multline}
	W=\sum\limits_h\int\limits_{\R^+}\frac{pdp}{2q(q^2+p^2+i\e)}T^{h_1h_2h^*}_{\bk_1,\bk_2;\bq,p}T^{h_3h_4h}_{\bk_3,\bk_4;-\bq,p}
	\\-\frac{S_2\pi^2}{8q^2}\Bigg(\int\limits_{0}^{\infty}\frac{dz_R}{e^{(k_3+k_4)z_R}}\bigg[\frac{8 \sqrt{\pi } (k_1+k_2)^{5/2}}{15 z^{9/2}}\\+\frac{E_{\frac{7}{2}}((k_1+k_2)
		z)-E_{\frac{9}{2}}((k_1+k_2) z)}{z^7}\bigg]\Bigg)
\end{multline}
with which \equref{eq: soft limit in unitary decomposition for four point} takes the rather nice form
\begin{multline}
	\lim\limits_{\bk_1\rightarrow 0} W=\sum\limits_h\int\limits_{\R^+}\frac{pdp}{2k_2(k_2^2+p^2+i\e)}T^{h_3h_4h}_{\bk_3,\bk_4;-\bq,p}\left(\lim\limits_{\bk_1\rightarrow 0} T^{h_1h_2h^*}_{\bk_1,\bk_2;\bq,p}\right)
	\\+W_{\text{AdS}}^{\text{soft}}
\end{multline}
where the intrinsically AdS contribution in this soft limit is immediately recognizable as the piece
\begin{multline}
	W_{\text{AdS}}^{\text{soft}}(k_2,k_3,k_4)=-\frac{S_2\pi^2}{8}\int\limits_{0}^{\infty}\frac{dz_R}{e^{(k_3+k_4)z_R}}\bigg[\frac{8 \sqrt{\pi k_2}}{15 z^{9/2}}\\+\frac{E_{\frac{7}{2}}(k_2
		z)-E_{\frac{9}{2}}(k_2 z)}{k_2^2z^7}\bigg]
\end{multline}
This expression by itself is of course divergent as one needs to choose a regularization scheme to extract explicit results; nevertheless, we can use this form for consistency checks, for instance we see that the expression obeys
\be 
W_{\text{AdS}}^{\text{soft}}(\l k_2, \l k_3, \l k_4)=\l^{4}W_{\text{AdS}}^{\text{soft}}(k_2,k_3,k_4)
\ee 
as dimensionally expected.

\subsection{Three point amplitudes}
\label{appendix: three point amplitudes}

Even though we have the technology to compute any $n-$point amplitude, the computation is tedious with lengthy results that provide little to no physical insight. The exception is the tree point amplitudes, and in this section, we are going to provide some explicit expressions for the three point amplitude $\cA$ of a AdS$_{d+1}$ particle dual to a conserved current of spin $l$; and then, we are going to analyze their soft limit. The explicit expression for $\cA$ reads as
	\begin{multline}
		\cA^{h_1,h_2,h_3}(\bk_1,\bk_2,\bk_3)=\left[\e_i^{h_1}(\bk_1)\e_j^{h_2}(\bk_2)\e_k^{h_3}(\bk_3)V^{ijk}(\bk_1,\bk_2,\bk_3)\right]\\\x\cA(\bk_1,\bk_2,\bk_3)
	\end{multline}
	for the scalar factor
\small 
\be 
{}&{}\cA(\bk_1,\bk_2,\bk_3)=\int\limits_{z=0}^\infty\frac{dz}{z^{d+1}}z^{4l}\left[\sqrt{\frac{2}{\pi}}z^{\frac{d}{2}-l}k_1^{\frac{d}{2}+l-2}K_{\frac{d}{2}+l-2}(k_1 z)\right]
\\{}&{}\x \left[\sqrt{\frac{2}{\pi}}z^{\frac{d}{2}-l}k_2^{\frac{d}{2}+l-2}K_{\frac{d}{2}+l-2}(k_2 z)\right]\left[\sqrt{\frac{2}{\pi}}z^{\frac{d}{2}-l}k_3^{\frac{d}{2}+l-2}K_{\frac{d}{2}+l-2}(k_3 z)\right]
\ee
\normalsize
where we choose an overall normalization coefficient $\sqrt{\frac{2}{\pi}}$ for later simplicity. We are also focusing on interactions which appear with $l$ derivatives in the Lagrangian, hence gauge theory and linearized gravity are $l=1$ and $l=2$ examples of it. This choice is the reason why we get $z^{4l}$ factor above \cite{Albayrak:2023kfk}.

We can compute the integral in a straightforward fashion, where some examples are provided in Table~\ref{table}.\footnote{Although we provide examples for gluons and gravitons, the same expressions are valid for higher spinning particles as well; for instance, the scalar factor of a spin$-3$ conserved current in AdS$_4$ is same as the graviton expression in AdS$_{6}$: we refer reader to \cite{Albayrak:2023kfk} for a discussion of such connections between web of theories.} We can verify the natural expectation that these expressions are non-singular in soft limit, i.e. $\bk_1\to 0$. In fact, somehow surprisingly, the soft limit does take a remarkably simple form:
\begin{multline}
\lim\limits_{k_1\to 0}\cA_3(k_1,k_2,k_3)=\frac{(2n-1)!!}{(-1)^n}\frac{k_2^{2n+1}-k_3^{2n+1}}{k_2^2-k_3^2}\\\text{ for }n=\frac{2l+d-5}{2}\in\Z_{\ge 0}
\end{multline}
We can further impose the momentum conservation to rewrite this in the expected power-law form of the propagator, i.e. $\lim\limits_{\bm\e\to 0}\cA_3(\e,k,\abs{-\bk-\bm\e})=(-1)^n(2n+1)!!k^{2n}$.

% Need to consolidate references

\bibliography{softref,collectiveReferenceLibrary}
\bibliographystyle{utphysModified}
\end{document}